\documentclass{article}%
\usepackage{amsmath}
\usepackage{amsfonts}
\usepackage{amssymb}
\usepackage{graphicx}%
\begin{document}

\title{ }

\textbf{Price equations with symmetric supply/demand; implications for fat
tails}

Carey Caginalp$^{1,2}$ and Gunduz Caginalp$^{1}$

October 14, 2018

\bigskip

\bigskip

$^{1}$ Mathematics Department, University of Pittsburgh, Pittsburgh, PA 15260

$^{2}$ Economic Science Institute, Chapman University, Orange, CA 92866

\bigskip

Email: CC \ carey\_caginalp@alumni.brown.edu, GC \ caginalp@pitt.edu

\bigskip

\textbf{Abstract.} Implementing a set of microeconomic criteria, we develop
price dynamics equations using a function of demand/supply with key symmetry
properties. The function of demand/supply can be linear or nonlinear. The type
of function determines the nature of the tail of the distribution based on the
randomness in the supply and demand. For example, if supply and demand are
normally distributed, and the function is assumed to be linear, then the
density of relative price change has behavior $x^{-2}$ for large $x$ (i.e.,
large deviations). The exponent approaches $-1$ if the function of supply and
demand involves a large exponent. The falloff is exponential, i.e., $e^{-x}$,
if the function of supply and demand is logarithmic.

\bigskip


\bigskip

\textbf{1. Introduction. }

\bigskip

Equations for price dynamics have generally fallen into one of two categories:
(a) continuum approaches for asset markets that assume infinite arbitrage and
stochastics without directly addressing supply and demand, (b) discrete
approaches that examine the micro-structure of supply and demand.

The latter approach that is used widely in the mathematical finance community
is expressed (see e.g., \cite{BA, BS, W}) in the continuum form as
\begin{equation}
P^{-1}dP=\mu dt+\sigma dW \label{cs}%
\end{equation}
where $P\left(  t\right)  $ is price at (continuous) time $t,$ while $W$ is
Brownian motion, and $\mu$ and $\sigma$ are the mean and standard deviation of
the stochastic process. This approach marginalizes the issues involving supply
and demand, modeling instead the price change as though $\left(
\ref{cs}\right)  $ were an empirically observed phenomenon. While there is
some empirical justification for this equation$,$ there is large discrepancy
between the implications for the frequency of unusual events \cite{F, KE, M,
MH, NG, NA2, XA, GGP, DJ, KH}. \ In particular, if one measures $\sigma$ for
the S\&P 500 then $\left(  \ref{cs}\right)  $ would imply that the frequency
of a 4\% drop, for example, occurs about one in millions of days, instead of
about 500 days, the observed frequency. This is a practical implication of the
puzzle known as "fat tails" that refers to rare events occurring much more
frequently than one might expect from classical results. More precisely, the
density of relative price changes is observed to fall as a power law rather
than exponentially.

The theoretical justification for equation $\left(  \ref{cs}\right)  $ is also
limited, and its widespread use is largely attributable to mathematical
convenience \cite{CH, CC}.

On the other hand, the approach developed by economists, i.e., (b), often
called excess demand, is expressed as%
\begin{equation}
p_{t}-p_{t-1}=d-s, \label{d}%
\end{equation}
\cite{WG, HQ, HGH, SC, PP}, with $p_{t}$ as the price at discrete time, $t$,
with supply, $s,$ and demand, $d$, at time $t-1.$ Equation $\left(
\ref{d}\right)  $ must be regarded as a local equation that describes change
at a particular set of values of $d$ and $s.$ Clearly, the price change will
depend upon the magnitudes of $d$ and $s$, and not just their differences. One
can remedy this feature by normalizing $d-s$ by $s,$ so that the right hand
side of $\left(  \ref{d}\right)  $ is $\left(  d-s\right)  /s$. Similarly, the
left hand side of $\left(  \ref{d}\right)  $ needs to be normalized, for
example by dividing by $p_{t-1}.$

A third approach to price dynamics was built on this perspective to model an
actively traded asset or commodity (see e.g. \cite{CB}). With active trading
one can regard the buy/sell orders as flow. This led to the asset flow
equations that were written in continuum form in 1990 (see \cite{CB}, and more
recent works, e.g., \cite{MA1} and references therein). The price equation has
the form%
\begin{equation}
\tau_{0}\frac{1}{P\left(  t\right)  }\frac{dP\left(  t\right)  }{dt}%
=\frac{D\left(  t\right)  -S\left(  t\right)  }{S\left(  t\right)  },
\label{dpc}%
\end{equation}
Here, $\tau_{0}$ is a time constant that also incorporates a constant rate
factor that can be placed in the right hand side. The difference in the two
approaches is due to fact that $\left(  \ref{cs}\right)  $ assumes infinite
arbitrage. This means that there is always capital that can take advantage of
mispricing of assets. In this way the deviation from realistic value will be
small and random.

\bigskip

\textbf{2. A General Symmetric Model.} Equation $\left(  \ref{dpc}\right)  $
is of course a linearization (in $D/S$) since relative price change may depend
nonlinearly on normalized excess demand. Another feature of the right hand
side is that it is not symmetric with respect to supply and demand. This is
not significant when supply and demand are approximately equal. However, as
$D\rightarrow0$ (with $S$ fixed) we see that the right hand side approaches
$-\infty$ but as $S\rightarrow0$ (with $D$ fixed) the right hand side
approaches $1.$

\bigskip

\textbf{2.1} \emph{Basic requirements} \ 

\bigskip

One way to impose symmetry between $D$ and $S$ is to write in place of
$\left(  \ref{dpc}\right)  $ the equation%
\begin{equation}
\tau_{0}\frac{1}{P\left(  t\right)  }\frac{dP\left(  t\right)  }{dt}=\frac
{1}{2}\left(  \frac{D}{S}-\frac{S}{D}\right)  . \label{dps}%
\end{equation}

Note that the two equations $\left(  \ref{dps}\right)  $ and $\left(
\ref{dpc}\right)  $ have the same value for the first term in the perturbation
of $D=1+\delta$ and $S=1+\varepsilon$ about $\delta=\varepsilon=0$. Equation
$\left(  \ref{dps}\right)  $ is a basic model that satisfies a number of
requirements for a price equation: $\left(  i\right)  $ The price derivative
vanishes when $D=S$ so that price does not change in equilibrium. When $D>S,$
prices rise, and vice-versa. $\left(  ii\right)  $ The roles of $D$ and $S$
are anti-symmetric, in the sense that $D/S-S/D=-\left(  S/D-D/S\right)  $.
$\left(  iii\right)  $ A small change in the positive direction for supply,
$S$, has the same effect as a small change in the negative direction for
demand, $D.$ $(iv)$ When $D\rightarrow\infty$ (with $S$ fixed) the relative
price change diverges to $\infty;$ when $S\rightarrow\infty$ (with $D$ fixed)
it diverges to $-\infty.$

The equation $\left(  \ref{dps}\right)  $ is a simple prototype exhibiting the
features required for a price adjustment equation. We can consider a more
general form by stipulating the requirements for a function $G:\mathbb{R}%
^{+}\mathbb{\rightarrow}\mathbb{R}$ so $G\left(  D/S\right)  $ replaces the
right hand side of $\left(  \ref{dps}\right)  ,$ i.e.,%
\begin{equation}
\frac{d\log P\left(  t\right)  }{dt}=G\left(  \frac{D\left(  t\right)
}{S\left(  t\right)  }\right)  \label{p}%
\end{equation}

\bigskip

\bigskip

\textbf{2.2} \emph{Condition} $G$

\bigskip

The function $G:\mathbb{R}^{+}\rightarrow\mathbb{R}$ is required to be a twice
differentiable function satisfying the following:

$\left(  i\right)  $ $G\left(  1\right)  =0,$\ $\left(  ii\right)  $
$G^{\prime}\left(  x\right)  >0$ \ all $x\in\mathbb{R}^{+},\ \left(
iii\right)  \ G\left(  x\right)  =-G\left(  \frac{1}{x}\right)  ,\ $%
\[
\left(  iv\right)  \ \ \lim_{x\rightarrow\infty}xG^{\prime}\left(  x\right)
=\infty\ \ \text{and}\ \ \ \lim_{x\rightarrow0+}xG^{\prime}\left(  x\right)
=\infty.
\]

\[
\left(  v\right)  \ \left(  xG^{\prime}\left(  x\right)  \right)  ^{\prime
}\ \ is\ \ \left\{
\begin{array}
[c]{ccc}%
<0 & if & x<1\\
>0 & if & x>1
\end{array}
\right.  .\ \ ///
\]
These properties imply the following:%
\begin{equation}
xG^{\prime}\left(  x\right)  =\frac{1}{x}G^{\prime}\left(  \frac{1}{x}\right)
. \label{deriv}%
\end{equation}%
\begin{equation}
\lim_{x\rightarrow\infty}G\left(  x\right)  =\infty. \label{inf}%
\end{equation}
The first of these follows from differentiating $\left(  iii\right)  $. To
prove $\left(  \ref{inf}\right)  $ observe that\ with $C:=G^{\prime}\left(
1\right)  >0$ and $x>1,$ condition $\left(  iv\right)  $ implies%
\[
G^{\prime}\left(  x\right)  >Cx^{-1}\ if\ x>1.
\]
Integrating, we obtain, since $G\left(  1\right)  =0,$%
\[
G\left(  x\right)  \geq C\int_{1}^{x}s^{-1}ds=C\log x.
\]
and so $G\left(  x\right)  \rightarrow\infty$ as $x\rightarrow\infty$.

Note that since $G^{\prime}\left(  x\right)  >0,$ we can always normalize so
that $G^{\prime}\left(  1\right)  =1$ and incorporate the constant into the
time variable in the price equation $\left(  \ref{p}\right)  $.

Conditions $\left(  i\right)  -\left(  iii\right)  $ are basic requirements
for a symmetric price function, while $\left(  iv\right)  $ and $\left(
v\right)  $ are useful symmetry properties for construction of stochastic equations.

\bigskip

\textbf{2.3 }\emph{Examples of functions that satisfy Condition G}

\bigskip

In addition to the function in $\left(  \ref{dps}\right)  $ one can readily
verify that the following functions also satisfy this condition:

$\left(  i\right)  $ $G\left(  x\right)  =x^{q}-x^{-q}$ for $q>0;$

$\left(  ii\right)  $ $G\left(  x\right)  =\left(  x-x^{-1}\right)  ^{q}$ for
$q$ an odd positive integer.

\bigskip

\textbf{3. Fat Tails and Demand, Supply Quotient.} The price dynamics
equations $\left(  \ref{dpc}\right)  $ and $\left(  \ref{dps}\right)  $ both
involve the quotient $D/S$. It is reasonable to assume, based on the Central
Limit Theorem\textbf{, }that given many agents placing buy and sell orders
into the market, the distribution of orders at the market price will be normal
(Gaussian). The question of the tail of the distribution then entails the
study of a quotient of normals (see, e.g., \cite{GE, DA, DR, MA, TO}).
Generally, we expect that $S$ and $D$ will have a negative correlation, and in
an idealized setting, they will have correlation $-1$ as random events that
increase supply tend to decrease demand. Earlier work \cite{CC} on this issue
using $\left(  \ref{dps}\right)  $ has produced the result that if $D$ and $S$
are described by a bivariate normal distribution, the density, $f\left(
x\right)  $, will falloff with exponent $-2,$ i.e., $\ f\left(  x\right)  \sim
x^{-2}$ for large $x.$ Moreover, a very simple formula was found for the
density in the special case when the correlation between $D$ and $S$ is $-1$
(anti-correlation). A key theorem proved in \cite{CC}, on which subsequent
results will be based, is stated below.

\bigskip

\textbf{3.1 }\emph{Quotient of Normals }

\bigskip

\textbf{Theorem.} If $R:=D/S$ where $D$ and $S$ are bivariate normal random
variables with strictly positive means and variances, $\mu_{1},\mu_{2}%
,\sigma_{1},\sigma_{2},$ and correlation $-1<\rho<1,$ then the density of $R,$
falls off as%
\[
f_{R}\left(  x\right)  \sim f_{0}x^{-2},
\]
where $f_{0}$ depends on $\mu_{1},\mu_{2},\sigma_{1},\sigma_{2}$ and $\rho.$

For $\rho=-1$ one has the exact expression (for $x\not =-\sigma_{1}/\sigma
_{2}$)%
\[
f_{R}\left(  x\right)  =\frac{\mu_{1}\sigma_{2}+\mu_{2}\sigma_{1}}{\sqrt{2\pi
}}\frac{e^{-\frac{1}{2}\left(  \frac{\mu_{2}x-\mu_{1}}{\sigma_{2}x+\sigma_{1}%
}\right)  ^{2}}}{\left(  \sigma_{2}x+\sigma_{1}\right)  ^{2}}%
\]
and $f_{R}\left(  -\sigma_{1}/\sigma_{2}\right)  =0.$

\bigskip

\bigskip

\textbf{3.2} \emph{Tail behavior}

\bigskip

We show below that the $f\left(  x\right)  \sim x^{-2}$ behavior is also valid
for the symmetric model $\left(  \ref{dps}\right)  .$

\bigskip

\textbf{Lemma.} Let $R$ be a random variable with density $f_{R}$ such that
$f_{R}\left(  x\right)  \sim x^{-p}$ with $p>1$ for large $x.$ For $q>0$ (so
that $1-p-q<0$) one has:

$\left(  a\right)  $ $f_{R^{q}}\left(  x\right)  \sim x^{\frac{1-p}{q}-1},$
$\left(  b\right)  $ $f_{R^{q}-R^{-q}}\left(  x\right)  \sim x^{\frac{1-p}%
{q}-1}$ and

$\left(  c\right)  $ $f_{\left(  R-R^{-1}\right)  ^{q}}\left(  x\right)  \sim
x^{\frac{1-p}{q}-1}$ .

Proof. $\left(  a\right)  $ We note $P\left\{  R^{q}\leq x,R>0\right\}
=P\left\{  R\leq x^{1/q},R>0\right\}  =\int_{0}^{x^{1/q}}f\left(  s\right)
ds.$

The density for $R^{q},$ denoted $f_{R^{q}},$ is given by
\begin{align*}
f_{R^{q}}\left(  x\right)   &  =\partial_{x}P\left\{  R^{q}\leq x,R>0\right\}
,\ i.e.,\\
f_{R^{q}}\left(  x\right)   &  =\partial_{x}F\left(  x^{1/q}\right)
=F^{\prime}\left(  x^{1/q}\right)  \frac{1}{q}x^{1/q-1}\\
&  =f\left(  x^{1/q}\right)  \frac{1}{q}x^{1/q-1}\sim\left(  x^{1/q}\right)
^{-p}x^{1/q-1}=x^{\frac{1-p}{q}-1}.
\end{align*}

The remaining parts of the theorem can be obtained by similar methods. $///$

\bigskip

Note that the calculations are similar for $0<p<1$ (provided we impose
$1-p-q<0$), but the mean does not exist in this range.

\bigskip

\textbf{Theorem.} Let $D$ and $S$ be bivariate normals with correlation
$\rho<1,$ and let $R:=D/S.$ Then the density of the functions $G_{1}\left(
R\right)  =R^{q}-R^{-q}$ and $G_{2}\left(  R\right)  =\left(  R-1/R\right)
^{q}$ both satisfy the large $x$ behavior%
\[
f\left(  x\right)  \sim f_{0}x^{-1-1/q}.
\]

\bigskip

\textbf{Remark. }When $\rho=-1,$ we have the exact density for $f_{\left(
D/S\right)  ^{q}}$%

\begin{align*}
f_{\left(  D/S\right)  ^{q}}\left(  x\right)   &  =f\left(  x^{1/q}\right)
\frac{1}{q}x^{1/q-1}\\
&  =\frac{\mu_{1}\sigma_{2}+\mu_{2}\sigma_{1}}{\sqrt{2\pi}q}\frac{e^{-\frac
{1}{2}\left(  \frac{\mu_{2}x^{1/q}-\mu_{1}}{\sigma_{2}x^{1/q}+\sigma_{1}%
}\right)  ^{2}}}{\left(  \sigma_{2}x^{1/q}+\sigma_{1}\right)  ^{2}}x^{1/q-1}%
\end{align*}
which, of course, has the $x^{-1-1/q}$ decay.

\bigskip

\textbf{3.3} \emph{Limits}

\bigskip

Consider the case $R=D/S$ with $D$ and $S$ normal with arbitrary correlation
$\rho<1.$ As noted above, $f_{R}\left(  x\right)  $ $\sim x^{-2},$ the decay
for $R^{q}$ is $f_{R^{q}}\left(  x\right)  \sim x^{-1-1/q}.$ Recall that the
density for both $R^{q}-R^{-q}$ and $\left(  R-R^{-1}\right)  ^{q}$ falls off
with the same exponent as $R^{q}.$ \ Under these conditions we note the
following limits.

As $q\rightarrow\infty$ the decay goes to $x^{-1}.$ Note that large $q$ means
that a change in the demand/supply makes a larger change in relative price,
i.e., $P^{-1}dP/dt$. Hence, it appears that, for any correlation, $\rho$,
between $D$ and $S$ one has that as $q$ increases (i.e., prices are very
sensitive to supply/demand changes), the decay exponent moves closer to
$x^{-1}.$

As $q\rightarrow0,$ i.e., prices do not vary much as supply/demand changes, so
that the exponent of $x^{-1-1/q}$ diverges to $-\infty.$

\bigskip

\textbf{3.4 }\emph{Logarithmic functions}\ 

\bigskip

This last limit suggests that examining $\log\left(  D/S\right)  $ may yield
an exponential decay. I.e. we use the equation%
\[
P^{-1}\frac{dP}{dt}=\log\left(  D/S\right)  .
\]
Letting $R:=D/S,$ we know that if $D,S$ are normal, then $f_{R}\left(
x\right)  \sim x^{-2}\ $so that%

\[
P\left\{  \log R\leq x,R>0\right\}  =P\left\{  R\leq e^{x},R>0\right\}
=\int_{0}^{e^{x}}f\left(  s\right)  ds.
\]
Taking the derivative, we have then%

\begin{align*}
f_{\log R}\left(  x\right)   &  =\partial_{x}P\left\{  \log R\leq
x,R>0\right\} \\
&  =\partial_{x}\int_{0}^{e^{x}}f\left(  s\right)  ds=e^{x}f\left(
e^{x}\right)  \sim e^{x}\left(  e^{x}\right)  ^{-2}=e^{-x}.
\end{align*}

Hence, if $D$ and $S$ are bivariate normal with correlation less than $1$, and
the relative change in price is proportional to $\log D/S,$ then the relative
price change has a density that falls off as $e^{-x}.$ Also, if $p^{-1}$ is an
odd positive integer, then $G\left(  D/S\right)  =\left[  \log\left(
D/S\right)  \right]  ^{1/p}$ yields a decay of $x^{p-1}e^{-x^{p}}.$

Note that $G\left(  x\right)  =\left(  \log x\right)  ^{q}$, with $q$ an odd
integer greater than 1, satisfies Condition $G,$ while $G\left(  x\right)
=\log x$ satisfies only the conditions $\left(  i\right)  -\left(  iii\right)
.$ In place of $\left(  iv\right)  $ and $\left(  v\right)  $ it satisfies the
symmetry condition $xG^{\prime}\left(  x\right)  =1$ for all $x\in
\mathbb{R}^{+}.$

\bigskip

\textbf{4. Conclusion.}

\bigskip

The results above establish a link between the relative price changes and the
exponent of the fat tails through $\left(  \ref{p}\right)  .$ One of the
problems in empirically estimating the likelihood of rare events is that one
would need a very long time history, but this often takes us back to a
different time period that may be irrelevant. We can use our results to
estimate the exponents by examining a much smaller data set and fitting $G.$
To be precise, choose small $\delta t$ and $\Delta t$ so that
\[
\delta t<<\Delta t.
\]
We approximate the left hand side of $\left(  \ref{p}\right)  $ as
$P^{-1}\delta P/\delta t$ and obtain statistics for all intervals of $\delta
t$ within $\left(  t,t+\Delta t\right)  .$ Then we can determine the function
$G$ that best fits the data. Using the results of Sections 3.2 and 3.4 one can
then ascertain whether the decay in the density falls off exponentially (i.e.,
$G$ is a logarithmic function) or with fat tails (with a specific exponent).

\bigskip

\textbf{Acknowledgements.} The authors thank the Economic Science Institute
and the Hayek Foundation for their support. Comments by two anonymous referees
are also gratefully acknowledged.

\end{document}